\documentclass[letterpaper]{article}
\usepackage{graphicx}
\usepackage{epsfig}

\begin{document}
\date {}
\title{A CRYOGENIC UNDERGROUND OBSERVATORY FOR RARE EVENTS: CUORE, AN UPDATE }

\maketitle

\footnotetext[1]{Dipartimento di Fisica dell'Universit\`{a} di
Milano-Bicocca e Sezione di Milano dell'INFN, Milano I-20126,
Italy}

\footnotetext[2]{University of South Carolina, Dept.of Physics
and Astronomy, Columbia, South Carolina, USA 29208 }

\footnotetext[3]{Lawrence Berkeley National Laboratory, Berkeley,
California, 94720, USA}

\footnotetext[4]{Dept. of Materials Science and Mineral
Engineering, University of California, Berkeley, California
94720, USA }

\footnotetext[5]{Dipartimento di Fisica dell'Universit\`{a} di
Firenze e Sezione di Firenze dell'INFN, Firenze I-50125, Italy }

\footnotetext[6]{Laboratori Nazionali del Gran Sasso, I-67010,
Assergi (L'Aquila), Italy}

\footnotetext[7]{Laboratorio de Fisica Nuclear y Altas Energias,
Universid\`{a}d de Zasagoza, 50009 Zaragoza, Spain }

\footnotetext[8]{Kamerling Onnes Laboratory, Leiden University,
2300 RAQ, Leiden, The Netherlands }

\footnotetext[9]{Dipartimento di Scienze Chimiche, Fisiche e
Matematiche dell'Universit\`{a} dell'Insubria e Sezione di Milano
dell'INFN, Como I-22100, Italy }

\footnotetext[10]{Laboratori Nazionali de Legnaro, Via Romea 4,
I-35020 Legnaro ( Padova ), Italy }

\author{ A.Alessandrello\footnotemark[1], C.Arnaboldi \footnotemark[1]
,F.T.Avignone\footnotemark[2],
J.Beeman\footnotemark[3]$^,$\footnotemark[4],
M.Barucci\footnotemark[5], M.Balata\footnotemark[6],
C.Brofferio\footnotemark[1], C.Bucci\footnotemark[6],
S.Cebrian\footnotemark[7], R.J.Creswick\footnotemark[2],
S.Capelli\footnotemark[1], L.Carbone\footnotemark[1],
O.Cremonesi\footnotemark[1], A.de Ward\footnotemark[8],
E.Fiorini\footnotemark[1], H.A.Farach\footnotemark[2],
G.Frossati\footnotemark[8], A.Giuliani\footnotemark[9],
D.Giugni\footnotemark[1],
E.E.Haller\footnotemark[3]$^,$\footnotemark[4],
I.G.Irastorza\footnotemark[7], R.J.McDonald\footnotemark[3],
A.Morales\footnotemark[7], E.B.Norman\footnotemark[3],
P.Negri\footnotemark[1], A.Nucciotti\footnotemark[1],
M.Pedretti\footnotemark[9], C.Robes\footnotemark[6],
V.Palmieri\footnotemark[10], M.Pavan\footnotemark[1],
G.Pessina\footnotemark[1], S.Pirro\footnotemark[1],
E.Previtali\footnotemark[1], C.Rosenfeld\footnotemark[2],
A.R.Smith\footnotemark[3], M.Sisti\footnotemark[1],
G.Ventura\footnotemark[5], M.Vanzini\footnotemark[1],
and L.Zanotti \footnotemark[1]\\
\centerline{(The CUORE COLLABORATION)}}

%\address{}

\begin{abstract}
{CUORE is a proposed tightly packed array of 1000 $ TeO_{2} $
bolometers, each being a cube 5 cm on a side with a mass of 750
gms. The array consists of 25 vertical towers,  arranged in a
square, of 5 towers by 5 towers, each containing 10 layers of 4
crystals. The design of the detector is optimized for ultralow-
background searches for neutrinoless double beta decay of $
^{130}Te $ (33.8\% abundance), cold dark matter, solar axions,
and rare nuclear decays. A preliminary experiment involving 20
crystals of various sizes (MIBETA) has been completed, and a
single CUORE tower is being constructed as a smaller scale
experiment called CUORICINO. The expected performance and
sensitivity, based on Monte Carlo simulations and extrapolations
of present results, are reported. }
\end{abstract}

\section{\rm INTRODUCTION}

  Neutrinoless double-beta decay, is a process by which
  two neutrons in a nucleus  beta decay by exchanging a virtual
  Majorana neutrino, and each  emitting an electron.
  This violates lepton number conservation $(\Delta l = 2)$ \cite{furry}.
  There are many reviews on the subject [2 - 4].

  The decay rate for the process involving the exchange of a
  Majorana neutrino can be written as follows:
 \begin{equation}
 \lambda^{0\nu}_{\beta\beta} =
 G^{0\nu}(E_0,Z)< m_\nu >^{2}|M^{0\nu}_f-( g_A /g_V )^2M^{0\nu}_{GT}|^{2}.
\end{equation}
In equation (1) $ G^{0\nu}$ is the two-body phase-space factor
including coupling constants, $ M^{0\nu}_f $ and $ M^{0\nu}_{GT} $
are the Fermi and Gamow-Teller nuclear matrix elements
respectively, and $g_A$ and $g_V$ are the axial-vector and vector
relative weak coupling constants, respectively. The quantity $ <
m_\nu > $ is the effective Majorana neutrino mass given by:
\begin{equation}
< m_\nu > \equiv \vert \sum^{2n}_{k=1} \lambda^{cp}_k (
U^L_{lk})^{2} m_k \vert ,
\end{equation}
where $\lambda^{CP}_k$ is the $CP$ eigenvalue associated with the
$ k^{th} $ neutrino mass eigenstate  ($ \pm 1$ for $ CP$
conservation), $ U^L_{lk} $ is the $ (l,k )$ matrix element of the
transformation between flavor eigenstates $ | \nu_l> $ and mass
eigenstates $ |\nu_k> $ for left handed neutrinos;
\begin{equation}
|\nu_l > = \sum U^L_{lk} | \nu_k > ,
\end{equation}
and $ m_k $ is the mass of the $ k^{th}$ neutrino mass eigenstate.

The effective Majorana neutrino mass, $ < m_\nu > $, is directly
derivable from the measured half-life of the decay as follows:
\begin{equation}
< m_\nu > = m_e ( F_N T^{ 0\nu}_{ 1/2 } )^{ -1/2 } eV,
\end{equation}
where $ F_N \equiv G^{ 0\nu } | M^{ 0\nu}_f - ( g_A / g_V ) M^{
0\nu }_{ G T } |^{2}$, and $ m_e $ is the electron mass. This
quantity derives from nuclear structure calculations and is model
dependent as shown later.

     The most sensitive experiments thus far utilize germanium
     detectors isotopically enriched in $ ^{76}Ge $ from $ 7.78\% $
     abundance to $ \sim 86\% $. This activity began with natural
     abundance $ Ge $ detectors by Fiorini et al; in Milan \cite{fiorini}
     evolving over the years to the first experiments with small
     isotopically enriched $Ge$ detectors \cite{vasenko}, and finally to
     the two present multi-kilogram isotopically enriched
     $^{76}Ge$ experiments: Heidelberg Moscow \cite{bandis} and  IGEX
     \cite{aalseth}. These experiments have achieved lower bounds on the
     half-life of the decay $ ^{76}Ge \rightarrow ^{76}Se $
     $+ 2e^- : T^{0\nu} _{1/2} > 1.9 \times 10^{25}y $ \cite{bandis} and
     $ T^{0\nu} _{1/2} > 1.6 \times 10^{25}y $ \cite  {aalseth}. Reference
     \cite{bandis} has about four times the exposure as reference \cite{aalseth}
     with data of similar quality. This strongly implies that these
     experiments with the order of $100$ moles of $^{76}Ge$
     each are have reached their point of diminishing returns.
     Their continuation will yield little more information of
     fundamental interest. The latest large-space shell model
     calculation yields $ F_N = 1.41 \times 10^{-14}y^{-1}$\cite{cauries} .
     This value implies that the above half-lives yield
     $ < m _\nu > \leq 1.0\; eV $. Other calculations, discussed later,
     yield values as small as $ 0.3\; eV $.

     Where should the field of $ \beta\beta $ - decay go from
     here? Suppose we consider the observed neutrino oscillations
     in the data from atmospheric neutrinos \cite{fukuda} and solar
     neutrinos \cite{hampel}. Considering these data, what probable
     range of $ < m_\nu > $ is implied? Would it be large enough
     for a direct observation of $ 0\nu\beta\beta $ decay? If so,
     what technique would be the best for a possible discovery
     experiment? How much would such an experiment cost? We will
     address these questions in an effort to demonstrate that  CUORE,
      an array of $1000$, $750\; gm$ $TeO_2$ bolometers, is the
     best approach presently available . It can be launched without isotopic
     enrichment nor extensive  R \& D , and that it
     can achieve next generation sensitivity.

     \section{\rm THEORETICAL MOTIVATION: PROBABLE \\ NEUTRINO SCENARIOS}

     The Superkamiokande data imply maximal mixing of $ \nu_\mu $
     with $ \nu_\tau $ with $ \delta m^2 _{23} \simeq 3\times10^{-3}\; eV^2
     $. The solar neutrino data from SuperK and from SNO also imply that
     the small mixing angle solution to the solar neutrino problem
     is disfavored, so that $ \delta m^2 $ (solar) $ \simeq (
     10^{-5}-10^{-4})\; eV^2 $. Based on these interpretations, one
     probable scenario for the neutrino mixing matrix has
      the following approximate form:

\begin{equation}
{\Bigg( \begin{array}{c}
  {\nu_e} \\
  {\nu_{\mu}} \\
  {\nu_{\tau}}
\end{array}\Bigg) } \; = \;
{\Bigg(
\begin{array}{ccc}
  1/\sqrt{2} & 1/\sqrt{2} & 0 \\
  -1/2 & 1/2 & 1/\sqrt{2} \\
  1/2 & -1/2 & 1/\sqrt{2}
\end{array}
\Bigg)} {\Bigg( \begin{array}{c}
  {\nu_1} \\
  {\nu_{2}} \\
  {\nu_{3}}
\end{array}\Bigg) }
\end{equation}

     The neutrino masses can be arranged in two hierarchical
     patterns in which $ \delta m^2_{31} \simeq \delta m^2_{32}$
      $\sim 3 \times 10^{-3}\; eV^2 $, and $ \delta m^2_{21} \sim
      ( 10^{-5} - 10^{-4})\; eV^2$. With the available data, it is
      not possible to determine which hierarchy, $ m_3>m_1 (m_2)$,
      or $ m_1(m_2) > m_3$, is the correct one, nor do we know the
      absolute value of any of the mass eigenstates.

      The consideration of reactor neutrino and atmospheric
      neutrino data together strongly implies that the atmospheric
      neutrino oscillations are very dominantly $ \nu_\mu \rightarrow $
      $\nu_\tau (\nu_\mu \rightarrow \nu_\tau )$, which implies, as seen
      from equation \cite{fiorini}, that $ \nu_e $ is very dominantly a mixture of
      $ \nu_1 $ and $ \nu_2 $. In this case there will be one
      relative $ CP $ phase, $ \epsilon $, and equation reduces
      to the approximate form:
\begin{equation}
< m_\nu > = \frac{1}{2} ( m_1 + \epsilon m_2 ),
\end{equation}
where we recall that the large mixing angle solution of the solar
neutrino problem implies
\begin{equation}
( m^2_2 - m^2_1 ) = ( 10^{-5} - 10^{-4})\; eV^2.
\end{equation}
This yields four cases to be analyzed: $ (a) m_1 \simeq 0 $, $(b)$
$ m_1 >> 0.01\; eV $, $ (c) m_3 \simeq 0 $, and $ (d)$ the
existence of a mass scale, $ M $, where $ M >> 0.055\; eV $.

\begin{description}
  \item[a] If $m_1=0$, $m_2=(0.003-0.01)\;eV$ and
$<m_{\nu}>=\frac{m_2}{2}$.
  \item[b] If $m_1>>0.01\;eV\equiv M$. $<m_{\nu}>\simeq\frac{M}{2}(1+\epsilon )=0$ or
$M$.
  \item[c] If $m_3=0$, $m_1\simeq m_2\simeq 0.055\;eV$. $<m_2>\simeq 0$ or $0.055\; eV$.
  \item[d] If $M>>0.055\;eV$, $m_1\simeq m_2\simeq
M+0.055\;eV$. $<m_{\nu}>\simeq\frac{m_1}{2}(1+\epsilon )$.
\end{description}

If we assume then that $ \epsilon \simeq + 1 $, and that
neutrinos are Majorana particles, then it is very probable that
$< m_{\nu} > $ lies between $ 0.01 eV $ and the present bound from
$^{76}Ge$ experiments.

The requirements for a next generation experiment can easily be
deduced by reference to equation (8).
\begin{equation}
T^{0\nu}_{1/2} = {(\ln{2})Nt\over{c}},
\end{equation}
where $ N $ is the number of parent nuclei, $ t $ is the counting
time, and $ c $ is the total number of counts, dominantly
background. To improve the sensitivity to $ < m_\nu >$ by a
factor of $ 10^{-2}$ from the present $ 1\; eV $ to $ 0.01\; eV $,
one must increase the quantity $ Nt / c $ by a factor of $ 10^4$.
The quantity $ N $ can feasibly be increased by a factor of $
\sim 10^2$, over present experiments so that $ t/c $ must also be
improved by that amount. Since the present counting times are
probably about a factor of 5 less than a practical counting time,
the background should be reduced by a factor of between 10 and 20
below present levels. These are approximately the target
parameters of the next generation neutrinoless double-beta decay
experiments.

Georgi and Glashow give further motivation for more sensitive
next generation double - beta decay experiments \cite{howard}.
They discuss six "facts" deduced from atmospheric neutrino
experiments, and from solar neutrino experiments, and the
constraints imposed by the reactor neutrino experiments. They
conclude that if neutrinos play an essential role in the large
structure of the universe, their six "facts" are "mutually
consistent if and only if solar neutrino oscillations are nearly
maximal". They further state that stronger bounds on $
0\nu\beta\beta $ - decay could possibly constrain solar neutrino
data to allow only the just - so solution.

If, on the other hand, the small angle MSW solution somehow had
been the correct one, next generation $ 0\nu\beta\beta $ - decay
experiments could "exclude the cosmological relevance of relic
neutrinos" \cite{howard}.

\section{\rm PROPOSED NEXT GENERATION EXPERIMENTS}

There are six large volume experimental proposals in various
stages of development. CUORE will be discussed in detail later.
The remaining five in alphabetical order are: CAMEO, EXO, GENIUS,
MAJORANA, and MOON.

The $ CAMEO $ proposal would place enriched parent isotopes in and
near the center of the BOREXINO detector and Counting Test
Facility $ ( CTF) $ \cite{bellini}.

The proposed $ EXO $ detector would be either a large high
pressure $ ^{136}Xe $ gas Time Projection Chamber ( TPC ) or a
liquid TPC. It would contain tons of $ Xe $ isotopically enriched
in $ ^{136}Xe $ \cite{danilov}.

The $ GENIUS $ proposal involves between $1$ and $10\; tons$ of
"naked" germanium detectors, isotopically enriched to $86  \% $
in $ ^{76} Ge $, directly submerged in a large tank of liquid
nitrogen as a "clean"  shield \cite{klapdor}.

The $ Majorana $ proposal is a significant expansion of the IGEX
experiment with new segmented detectors, in a highly dense -
packed configuration and new pulse shape discrimination techniques
developed by PNNL and USC. The proposal involves the production
of 250, $2 \; kg $ isotopically enriched $ (86\%)\; Ge $
detectors, each segmented into 12 electrically independent
segments.

The Molybdenum Observatory of Neutrinos ( MOON ) proposal is a
major extension of the ELEGANTS detector. It involves between 1
and 3 tons of molybdenum foils isotropically enriched to 85\% in
$ ^{100}Mo $ inserted between plastic scintillators.

All of these experiments will require significant time,  R \& D,
and funding for isotopic enrichment as well as the development of
new techniques. The CUORE experiment on the other hand requires
no isotopic enrichment because the natural abundance of $
^{130}Te $ is $ (33.80 \pm 0.01)\% $, and the technique has
already been developed. A preliminary experiment, MIBETA, has
already been completed \cite{alessandrello}. In addition, a
preliminary trial experiment, CUORICINO, is being constructed at
this time \cite{aless}. It is one of the 25 towers of 40 of the
1000, $750 \; {gm}$ $Te0_{2} $ bolometers, which is a slight
change in the configuration initially designed \cite{aless}.
CUORICINO will contain $ 8.11\; {kg}$ of $ ^{130}Te $. The most
conservative nuclear structure calculations imply that $ ^{130}Te
$ is 2 times more effective in $ <m_\nu> $ sensitivity than $
^{76}Ge $, so that CUORICINO will be equal to at least $16.22 \;
kg $ of $ Ge $ enriched to 86\% in $ ^{76}Ge $. CUORE would be
equivalent to $407 \; kg $ of 86\% $ ^{76}Ge $ with the most
conservative nuclear matrix elements or 957 $ kg $ of 86\% $
^{76}Ge $ according to the largest theoretical matrix elements.
There are five nuclear structure calculations presented in Table
1 below [19 - 23].\\
\\

 TABLE 1: Theoretical values of $F_N$ for the double-beta decay of
 $^{76}Ge$ and $^{130}Te$ computed with five nuclear models.
\begin{center}
\begin{tabular}{|c|c|c|c|c|c|}
  % after \\: \hline or \cline{col1-col2} \cline{col3-col4} ...
  \hline
  $^{76}Ge$ & $^{130}Te$ &   &   &   &   \\ \hline
  $F_N(years)^{-1}$ & $F_N(years)^{-1}$ & $R(t)^+$ & $R(\epsilon )^*$ & model & Ref.
  \\ \hline
  $1.54\times 10^{-13}$ & $1.63\times 10^{-12}$ & 10.6 & 3.3 & Shell Model & [19] \\
  $1.14\times 10^{-13}$ & $1.08\times 10^{-12}$ & 9.6 & 3.1 & Generalized Seniority & [20] \\
  $1.86\times 10^{-14}$ & $3.96\times 10^{-13}$ & 21.8 & 4.7 & QRPA & [21] \\
  $1.24\times 10^{-13}$ & $4.98\times 10^{-13}$ & 3.9 & 2.0 & QRPA & [22] \\
  $1.14\times 10^{-13}$ & $5.33\times 10^{-13}$ & 4.7 & 2.2 & QRPA & [23] \\ \hline
\end{tabular}
\end{center}

\vskip 0.05cm
$^+$ $R(t)$ the ratio of $F_N(^{130}Te)/F_N(^{76}Ge)$. \\
$^*$ $R(\epsilon)$ The ratio $\sqrt{R(t)}$; the relative
sensitivity to $<m_{\nu}>$ of $^{130}Te$ to that of $^{76}Ge$. \\

\begin{figure}
\begin{center}
\mbox{\epsfig{file=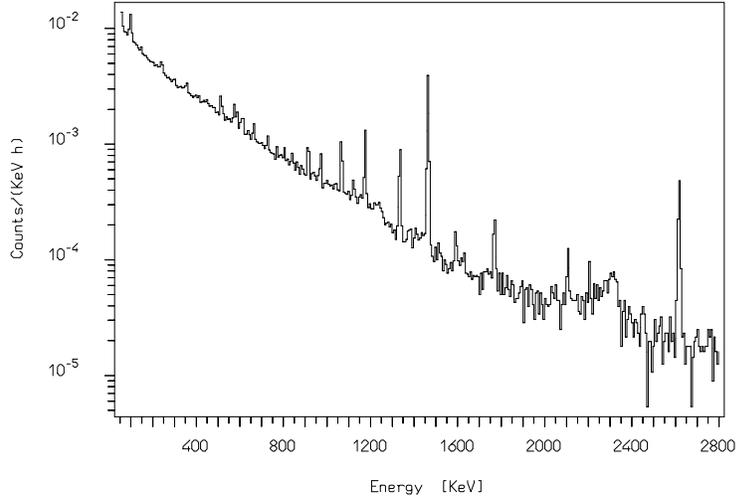,width=11cm}}
\linespread{.9}
\caption{\small{\em{Final spectrum of the MIBETA experiment.}}}
\linespread{1.1}
\end{center}
\end{figure}

The results of the preliminary experiment MIBETA reported by the
Milano - INFN group \cite{alessandrello} involved 20 natural $
TeO_{2}$ crystal bolometers  averaging $340 \; gm $ each for a
total mass of $6.8 \; kg $. This is equivalent to $1.84 \; kg $ of
$ ^{130}Te $. The array was run in a number of configurations with
various detectors operating at any one time as the array was used
for development of the technique. These experiments ran over a
total of 80613 hrs of operation but with several detector
configurations. The total exposure was $ Nt = 4.31 \times
10^{24}\;y $ with a bound on the number of counts in the $ 0 \nu
\beta \beta $ - decay region of $6.9^{+6.7}_{-5.9} $ for an upper
bound to 90\% $ CL $ of 17.96. The most recent bound on the half -
life is $ T^{0\nu}_{1/2} > 1.6 \times 10^{23}\;y $. This
corresponds to the following upper bounds on $ < m_\nu >$ for
eight nuclear structure calculations: $ 1.1eV $ \cite{haxton}, $
2.1 eV $ \cite{vogel}, $ 1.5eV $ \cite{engel}, $ 1.8\; eV $
\cite{staudt}, $ 2.4\; eV $ \cite{suhonen}, $ 1.9\; eV $
\cite{tomo}, $ 2.6\; eV $ \cite{faessler}, and $ 1.5\; eV $
\cite{barbero}. The range is from $1.1$ to $2.6 \; eV $, not much
larger than the conservative result from the $ ^{76}Ge $
experiments \cite{bandis,aalseth}.

The average background for this experiment was $0.5 counts /  kg
 /  keV  / y$; however, it was discovered later that the
crystals were polished with cerium  oxide which was measured in
the Gran Sasso Laboratory and found to be radioactive. In
addition there is clear evidence of neutron induced background.
We have conservatively estimated that the background can be
reduced by a factor of at least 10 in CUORICINO. The goal will be
to reduce the background significantly below this level, by a
factor of 40 or even more. \\

\break

TABLE 2: Projected sensitivities of CUORICINO depending on energy
resolution and background.

\begin{center}
\begin{tabular}{|c|c|c|c|c|}
  % after \\: \hline or \cline{col1-col2} \cline{col3-col4} ...
  \hline
  BKG  &  \multicolumn{2}{c}{ FWHM = 5 keV }\vline  &   \multicolumn{2}{c}{ FWHM = 2 keV}  \vline \\
  \cline{2-5}
   (c/keV kg y)& $\tau_{1/2}(y)$ & $<m_{\nu}> (eV)$ & $\tau_{1/2}(y)$ & $<m_{\nu}> (eV)$ \\
   \hline
  0.5 & $3.6\times 10^{24}\times t^{1/2}$ & $0.38\times t^{-1/4}$ & $5.7\times 10^{24}\times t^{1/2}$ & $0.30\times t^{-1/4}$ \\
  0.1 & $8.1\times 10^{24}\times t^{1/2}$ & $0.25\times t^{-1/4}$ & $1.3\times 10^{25}\times t^{1/2}$ & $0.20\times t^{-1/4}$ \\
  0.01& $2.6\times 10^{25}\times t^{1/2}$ & $0.14\times t^{-1/4}$ & $2.9\times 10^{25}\times t^{1/2}$ & $0.14\times t^{-1/4}$ \\ \hline
\end{tabular}
\end{center}

\vskip 0.3cm

The final spectrum of the MIBETA 20 crystal experiment reported in
reference \cite{alessandrello} is shown in Figure 1. While
improvements are being made at present in the background as well
as in the energy resolution, it is clear that this technology has
been clearly demonstrated by the results of the MIBETA
experiment, and will be further demonstrated by the CUORICINO
experiment being constructed at this time.

\section{\rm THE CUORE DETECTOR}

The CUORE will consist of an array of 1000 $ TeO_{2}$ bolometers
arranged in a square configuration of 25 towers of 40 crystal
each. The geometry of a single tower is shown in Figure 2a. A
sketch of CUORE is shown in Figure 2b.

\begin{figure}
\begin{center}
\mbox{\epsfig{file=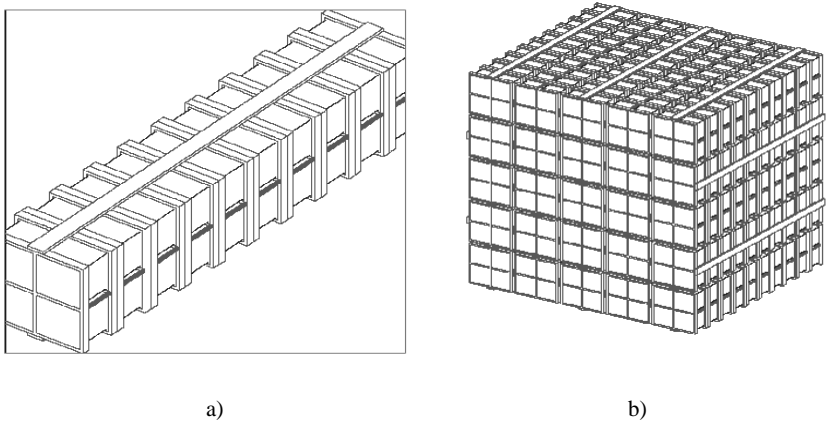,width=12cm}}
\linespread{.9}
\caption{\small{\em{a) A single tower of CUORE;   b) The CUORE detector.}}}
\linespread{1.1}
\end{center}
\end{figure}

The principle of operation of these bolometers is now well
understood \cite{twerenbold}. Telurium Oxide is a dielectric and
diamagnetic crystal, which when maintained at very low temperature
$ ( \sim 5 - 10^{\circ} millikelvin)$ has a very low heat
capacity. In fact the specific heat is proportional to the ratio
$ ( T / T_0 )^{3}$ where $T_0$ is the Debye temperature.
Accordingly, a very small energy absorbed by the crystal by a
nuclear decay or recoil by collision, can result in a measurable
increase in temperature. This temperature change can be recorded
using Neutron Transmutation Doped  (NTD)  germanium thermistors.
These devices were developed and produced by Haller at the
Lawrence Berkeley National Laboratory (LBNL) and UC Berkeley
Department of Material Science. These sensors have been made
unique in their uniformity of response and sensitivity by neutron
exposure control with neutron absorbing foils accompanying the
germanium in the reactor \cite{norman}.

The $ TeO_{2} $ crystals are produced by the Shanghai Quinhua
Material Company (SQM) in Shanghai, China. Crystals produced by
other organizations have proven to be inferior. A search of
potential suppliers in the U.S. revealed that the only dealers
found sold crystals produced by SQM or other companies outside of
the U.S.

Long periods of operation suffer small excursions in temperature
of the crystal array which deteriorates energy resolution. A
stabilization technique proven to be successful in the MIBETA 20
crystal array experiment will be employed. A periodic injection is
made of precisely known joule - power directly into the crystals
through heavily doped meanders in $Si$ chips glued to the surface
\cite{alessandrello}.

The single tower of 40 crystals is presently under construction.
It will be attached to the mixing chamber of the same dilution
refrigerator (DR) used in the MIBETA experiment
\cite{alessandrello} and run as a test. It will also be run as an
experiment called CUORICINO which is designed also to improve on
the present sensitivity to $ < m_\nu > $ obtained with
isotopically enriched $ Ge $ detectors \cite{bandis,aalseth}. By
the time significant funding from this proposal could be spent on
CUORE, CUORICINO will have proven the feasibility of the
extension of the MIBETA technology to large arrays. This, plus
the fact that CUORE requires no isotopic enrichment, puts CUORE
well ahead of all the other options of a truly next generation $ 0
\nu \beta \beta $ experiments. The technology, novel though it
is, is developed and to a large degree proven.

\section{\rm CONCLUSION}

The CUORE array will have $ 9.5 \times 10^{25} $ nuclei of $
^{130}Te $. If the background is conservatively reduced to $0.01
$ counts $ /keV /kg /yr $, then in one year of running, the
sensitivity of CUORE would be $ Te^{0\nu}_{1/2} > 1.1\times
10^{26}y $. This corresponds to $ < m_{\nu}> < 0.05 eV $. If
eventually, the background would be reduced to 0.001 counts $
/keV/ $ kg/y, the sensitivity with one year of counting would be $
T^{o\nu}_{1/2} > 3.6 \times 10^{26}y $, corresponding to $
<m_{\nu}> < 0.03eV $.

If in the two cases mentioned above, the detector was operated
for a decade, the bounds on $ <m_{\nu}> $ would be $ < 0.028 eV
$, and $ < 0.017 $ respectively.

If CUORE fulfills these expectations, it could be replicated by a
factory 6 for a similar cost ( conservatively speaking ) as any
of the experiments requiring isotopic enrichment.

The detector will also be used to search for cold dark matter
(CDM). The present thresholds of 5 $ keV $ are equivalent to 1.25
$ keV $ in ordinary $ Ge $ detectors, because ionization is 0.26
times as effective  in converting nuclear recoil energy into a
signal pulse as it is in converting photon energy. Such a large
array could efficiently search for a seasonal variation in the CDM
interaction rate. The large mass of $ Te $ theoretically enhances
the interaction rate of many CDM candidates.

The CUORE crystals will be placed in known crystalline
orientations which will allow a sensitive search for solar axions
using the technique introduced by Creswick et al.\cite{creswick},
and demonstrated by the experiment of Avignone et al.[32].

It should also be recognized that the highly granular
configuration of CUORE, equivalent to 10 layers of 100 crystal
each, approximately forms a cube with 512 crystals in an inner
cube with significant protection from a layer all around of 488
crystals. The coverage is not perfect because of necessary small
spaces between crystals; however, it will significantly reduce
background from gamma-rays coming from outside of the
configuration.

Finally, while the main emphasis is on building an array of $
TeO_{2} $ crystals, CUORE, or the CUORE technique can accommodate
any material that can be made into bolometers. The most promising
competing experiments are the two large proposed $ ^{76}Ge $
experiments GENIUS \cite{klapdor} and Majorana \cite{creswick}.
The direct observation of neutrinoless double beta decay
absolutely requires at least two different experiments with
different parent nuclei, if for no other reason than the
uncertainties in nuclear matrix elements. In $ ^{76}Ge $, for
example, this results in a factor of 4.3 in the value of
$<m_{\nu}>$, and a factor of 2.4 in the case of $ ^{130}Te $.
These uncertainties should be carefully considered when comparing
different proposals. For example, the present bounds on
$<m_{\nu}>$ from $ ^{76}Ge $ experiments range from 0.3 to 1.3 $
eV $, while they range from 1.1 to 2.6 $ eV $ from the small
MIBETA experiment recently completed [17]. CUORICINO should reach
a comparable sensitivity during its test period. Which bound would
be actually more restrictive? The answer lies in the
uncertainties in the nuclear physics.

\end{document}